# In search of the origin of Corona virus

**Rajasri Sen Jaiswal*, Richa Dobal, K. Thirumala Laksmhi and M. Siva**

Centre for Study on Rainfall and Radio wave Propagation, Sona College of Technology, Salem-5, Tamil Nadu, India

*For correspondence. (e-mail: rajasrisenjaiswal@gmail.com, rajasrisenjaiswal@sonatech.ac.in)

**Abstract-** In this paper, the authors aim to find out the origin of the corona and the duration of the present pandemic caused by it. Besides, they aim to find out the reason for such outbreaks occurring often in China. They make a hypothesis that the origin of the virus is embedded in the solar cycle. Next, they have proved the hypothesis by investigating the sunspot number, the cosmic ray flux data, the concentration of $^{10}$Be in the ice core in Greenland and based on the occurrence of past pandemics and endemics. The study shows that whenever the Sun is in the magnetically quiescent state, the world witnesses viral diseases. The study shows that the outbreak of the Corona is because of the minimum sunspot number during 2019-2020 and the Corona is likely to stay in the whole of 2020. The study indicates the occurrence of viral diseases in every 11-13 years. It also reveals that as the stratospheric thickness is the minimum at Wuhan, the virus had chosen this place as the first entry point to the Earth.

# Introduction

The world is kneeling under the grip of the pandemic caused by the Corona, a virus that had first erupted in China. All over the globe, many people have been infected by the deadly virus. The governments of all countries are trying to find out the cause of the occurrence of Corona and have initiated several safety measures. However, the number of infected people and the death toll continue to rise. In this paper, the authors have attempted to find out the cause of the outbreak. The very fact that the Sun is in the magnetically quiescent state in 2019- 2020 and the Corona outbreak is creating havoc since 2019, it had appeared to the corresponding author that the secret of the outbreak of the virus is embedded in the solar cycle. Thus, it is purely the author's hypothesis. To find out the cause of the Corona outbreak, in this paper, the authors have investigated the solar cycle from 1700-2019. Besides, they have investigated the cosmic ray flux data over Oulu, Finland, and Ahmedabad, India. Further, they have analyzed the concentration of $^{10}$Be collected in the ice core in Greenland. The authors have attempted, in particular, to find out when the Corona is likely to disappear, and when the next outbreak of viral disease will happen. Besides, they further aim to find out why the Corona has shown its first appearance in China.

   The Sun is the major driving force of the atmospheric processes. Sir William Herschel suggested a possible influence of the solar cycle on the Earth's atmosphere[1]. Several studies[2-4] have revealed the influence of the solar cycle on Earth. Studies show that the humidity[5] and temperature are strongly correlated to the sunspot number. Loon and Meehl[6] noticed that during the solar maxima, the sea surface temperature in the Pacific Ocean shows a pattern conducive to La Nina. The total solar irradiance (TSI), i.e. the incoming solar radiation at the top of the atmosphere, and the amount of solar radiation reaching the Earth's surface cause weather and climate on the Earth. The amount of radiation reaching a particular place on Earth is also governed by the Earth's orbital motion and its spin[7]. Thus, any changes in the atmospheric processes on the Earth depend on the modulations in solar radiation and the orbital and spinning motion of the Earth. Besides, the cosmic rays of galactic and solar origin also affect the atmospheric processes[8]. A Study[5] shows that the humidity at a place depends on the cosmic rays. The cosmic ray flux at a place shows strong correlations with the temperature over Oulu, Finland.

The total solar irradiance (TSI) depends on the magnetic activity of the Sun. It is found out that when the Sun is in the active state, then the TSI is higher[3] than when it is magnetically quiescent[9]. The sunspot number (SSN) is a manifestation of the level of activity of the Sun. When the Sun is magnetically active, then the SSN is high[10]. At this time, solar winds and solar flares are strong. Then, the cosmic rays, which are the high-energy protons and atomic nuclei[11] coming from the Sun and outside of the solar system, cannot penetrate the Earth's atmosphere as the strong magnetic field around the Sun reflects them into space. The Sun shows a periodicity in its cycle. Hathaway[12] has shown that SSN maxima occur after 11 years. When the Sun is magnetically quiescent, then the SSN is less[10]. At this time, the weak magnetic field around the Sun cannot shield the Earth from the high –energy cosmic rays[13]. It is because the energy of the cosmic rays is higher than the geomagnetic cut-off of the Earth. Then, the cosmic rays enter the Earth's atmosphere. The measured intensity of cosmic rays shows high during the magnetically quiescent state of the Sun[14]. The cosmic rays, which are primarily Hydrogen by 89%, Helium by 10% percent and all other elements by 1%[15], penetrate the Earth's atmosphere with the velocity of light. After reaching the Earth's atmosphere, the cosmic rays collide with the atoms of the Earth's upper atmosphere to produce secondary particles, viz., $^{14}C$ and $^{10}Be$[16]. The interaction of cosmic rays with the atmosphere produces ionization. The secondary particles produced during the ionization process by the interaction of the primary cosmic rays with the atoms present in the lower atmosphere form the ultra-fine aerosols. The aerosols act as cloud condensation nuclei. Thus, the cosmic rays affect the formation of clouds and rain. Observations[17,18] show that when the cosmic ray flux reaching the Earth decreases, the cloudiness and precipitation also decreases. The secondary particles further interact with the air molecules producing other particles, that shower towards the Earth's surface.

Apart from their impact on the Earth's climate, the cosmic rays have demonstrated the capability of inducing mutation in the viruses[19]. Mutations, i.e. the changes of the DNA replication process may be caused by several agents, of which the muons are the dominant components. Muons are produced by high energy cosmic rays[20] in the Earth's atmosphere, at an altitude of 15 km where high energy cosmic rays interact with the nuclei of atmospheric gases. These particles, due to their low interaction cross-section[21], travel towards the Earth with a velocity close to that of light. The muons penetrate deep underground and underwater, with the possibility of affecting biological entities[21]. A study by Ferrari and Szuszkiewicz[8] opines that the ionization caused by the cosmic rays might have caused the evolution of life on Earth. Rieseberg and Livingstone[19] have shown that the cosmic rays can change the chromosome in a reproductive cell. Mukherjee[22] opines that the cosmic rays may either induce mutation in the viruses already present in the air or the body of the human being. A study by Mukherjee[22] has shown that a heavy shower of cosmic ray flux, along with the sudden fall of electron flux during the last week of February 2009 to the first week of March 2009 was followed by the H1N1 outbreak in Mexico. Mukherjee[22] has reported continuous low electron flux before the outbreak of the disease. Between March and August of 2009, millions of people around the world were infected with influenza and respiratory failure[22]. Thus, the cosmic rays demonstrate to have the potential to cause mutation in living organisms. The research of Hope-Simpson[23] exhibits that the SSN cycle correlates with influenza pandemics. The influenza virus mutated to a different variety with the successive sunspot cycle[23].

In this paper, the authors aim to seek the reason for the Corona outbreak. By assuming that the origin of the virus lies in the mystery of the solar cycle, they have examined the correlation between the outbreak of viral diseases worldwide, and the sunspot number. The investigation is further complimented with the cosmic ray flux data and the production of $^{10}$Be over several years.

## Materials and methods

The investigation carried out in this paper is based on sunspot number data obtained from the Solar Influences Data Analysis Centre (SIDC), Royal Observatory of Belgium for the period 1700-2020[24]. The year of the outbreak of viral diseases has been obtained from the list of epidemics [25]. The cosmic ray flux data over Oulu have been obtained from the Cosmic Ray Station, Sodankyla Geophysical Observatory, the University of Oulu for the period 1964-2018[26]. The cosmic ray flux data over Ahmedabad during 1957-1958; 1964; and 1968-1973 have been obtained from the World Data Centre for Cosmic Ray, Physical Research Laboratory, Ahmedabad[27]. The data on $^{10}$Be are obtained from the NOAA National Centre for Environmental Information for the period 1700-1994[28].

In this paper, the time series of SSN, cosmic ray flux, and $^{10}$Be have been investigated. The simultaneous time series of any two parameters have been investigated for the period over which data for both the parameters are available. The analysis is based on the occurrence of the crests and troughs of the parameters. The correlation between the SSN and the cosmic ray flux over Oulu and Ahmedabad is based on a curve estimation technique using different models, viz. linear, power, logarithmic, logistics, inverse, inverse, cubic, compound, exponential, growth, quadratic, and sigmoid. The validity of the models is judged by the F test at a 5% level of significance.

**Results and discussion**

In an attempt to find out whether the outbreak of viral diseases always coincided or followed the low sunspot number, the occurrence of the two has been investigated during 1700-2020 (Table 1). Table 1 shows that the sunspot minima in the 11-13 year solar cycle have always brought viral diseases across the globe. For example, in 2009, the outbreak of swine flu worldwide followed the sunspot low in 2008. In 2009, the sunspot minimum was observed. The worldwide outbreak of plague in 1856-1860 was associated with a sunspot maximum in 1856. Table 1 further shows that at times, the outbreak of diseases also coincided with the sunspot maxima. For example, the outbreak of measles in 1788 was associated with the sunspot maxima in the same year. The outbreak of influenza worldwide in 1847- 1848 had seen the sunspot maximum in 1848.

Thus, a very low or a sunspot minimum is of concern as it is likely to cause an outbreak of viral diseases. The governments of all countries should be equipped with proper mitigation techniques after ever 11-13 years.

Figure 1 shows the variation of SSN during 1700- 2019, and $^{10}$Be for 1700-1994. Figure 1 shows that the SSN maxima and minima occur after a fixed interval. The maxima occur after 11.3 years and the minima occur after 11-13 years.

It is found out in Figure 1 that the $^{10}$Be shows its high after 2-9 years. The $^{10}$Be low also occurs after every 2-9 years. However, the minima of $^{10}$Be had occurred in 1872, 1939, 1958, 1980 and 1990. The values of $^{10}$Be in these years were 0.992, 0.931, 0.994, 0.956, and 0.851, respectively. Figure 1 further shows that the SSN minima are associated with $^{10}$Be maxima in the same year. At times, the $^{10}$Be maxima followed in the immediate next year when the SSN minima had occurred. For example, in 1700, 1711, 1766, 1784, 1798, 1810, 1823, 1878, 1923, 1933, 1944, the SSN minima and the $^{10}$Be peak had occurred. The $^{10}$Be peaks in 1734, 1756, 1776, 1857, 1890, 1914, 1965, and 1977 had followed the SSN troughs in the immediately previous years, implying that high $^{10}$Be concentration are associated with SSN minima or a very low SSN, i.e. during the magnetically quiescent state of the Sun. Sakurai *et al.*[29] also found anticorrelation between the sunspot number and the Be-7 concentration. However, it is noteworthy that rarely deviations are observed between the occurrence of the SSN troughs and $^{10}$Be peaks. For example, a $^{10}$Be peak was observed in 1842, but the SSN trough had happened in 1843. Besides, in 1986 the SSN trough was observed, and the $^{10}$Be was observed in 1988 and 1985. Also, in 1957, the $^{10}$Be peak coincided with the SSN peak.

Figure 2 *a* and *b* respectively shows the variation of sunspot number and cosmic ray flux over Oulu, Finland, and Ahmedabad, India. Figure 2 *a* shows that the SSN and cosmic ray flux maintain inverse relation over Olulu, i.e. the troughs of SSN time series fall on the crests of cosmic ray flux, and vice versa. At times, the cosmic ray flux maxima occur in the same year as the SSN minima. For example, in 1976 the maximum cosmic ray flux was associated with the minimum SSN. In 2000 and 2014, the SSN maxima coincided with the minimum cosmic ray flux. At times, the cosmic ray maxima followed the SSN minima. For example, the cosmic ray maxima of 1965, 1969, 1987, 1997, and 2009 respectively followed the SSN minima of 1964,

1968, 1986, 1996, and 2008. At times, the cosmic ray maxima follow the SSN maxima after two to three years. The cosmic ray maxima of 1982 and 1991 followed the SSN minima in 1979 and 1989, respectively. Figure 2 *b* shows that the SSN trough in 1973 coincided with the peak cosmic ray flux. Utomo[30] has shown that as the SSN increases, the cosmic ray flux decreases.

To find out the correlation between the SSN and cosmic ray flux, the authors have fitted the yearly SSN and cosmic ray flux data over Oulu and Ahmedabad to various models, viz. linear, power, logarithmic, logistic, inverse, cubic, quadratic, inverse, sigmoid, growth, exponential, and logistic. The validity of the models is judged by the F test at a 5% level of significance. The results of the investigation are described in Figure 3 *a* and *b*, respectively over Oulu and Ahmedabad. Figure 3 *a* and *b* shows that the SSN and cosmic ray flux bear a very significant relationship with an $R^2$ value of 0.747 over Oulu and 0.952 over Ahmedabad. The relationship over Oulu is found to be a quadratic one (Figure 3 *a*), while over Ahmedabad, a cubic relation (Figure 3 *b*) suits. Thus, Figure 3 *a* and *b* show that in the data-sparse regions, the cosmic ray flux can very well be estimated from the measured values of the SSN. However, the relationships are to be established using the data of cosmic ray flux and over a location and the SSN. Besides, the relationship is likely to vary from one location to another, depending on the latitude of the location.

Figure 4 *a* and *b* respectively represents the variations of cosmic ray flux and $^{10}$Be over Oulu Finland during 1964-1994 and Ahmedabad, India during 1968-1973. Figure 4 *a* shows that over Oulu, in 1965, both the cosmic ray flux and $^{10}$Be had attained the peak value. Similarly, in 1990 the minimum value of both cosmic rays and $^{10}$Be had happened. The $^{10}$Be maxima in 1977, and 1988 followed the cosmic ray maxima in 1976 and 1987, respectively. The cosmic ray minima in

1969, 1974, and 1982 were followed by the minima of $^{10}$Be in 1970, 1975, and 1984, respectively. In 1990, the minimum values of cosmic ray flux and $^{10}$Be had coincided.

However, some anomalies are observed in the occurrence of the crests and troughs of cosmic ray and $^{10}$Be time series. For example, in 1972, the peak $^{10}$Be was observed. The peak cosmic ray flux had occurred in 1973. Figure 4 *b* shows that the peak cosmic ray flux over Ahmedabad in 1973 was associated with a very high value of $^{10}$Be. However, the peak $^{10}$Be had occurred in 1972- one year before the cosmic ray peak had happened.

Thus, it appears that an increase in cosmic ray flux increases the production of $^{10}$Be. This further implies that when the Sun is in the magnetically quiescent state, marked by SSN minima, a large amount of cosmic ray flux penetrates the Earth's atmosphere and a large amount of $^{10}$Be deposit is observed. During 1690- 1700, the sunspot number was minimum. During this period, the $^{10}$Be showed the largest peak[31], implying that the cosmic ray flux reaching the Earth was high.

To find out the possibility of occurrence of the Corona to occur in 2019-2020, and to investigate the time duration of the existence of the same, the authors wish to find out the recent 20-year trend of the variations of SSN and cosmic ray flux during 1964- 2018. The period 1964-2018 has been chosen as both SSN and cosmic ray flux data are available for this period. Figure 5 *a* and *b* respectively shows the variations of SSN and cosmic ray flux over Oulu during 1979-1998 and 1999-2018. Figure 5 *a* and *b* shows that in the recent 20-year periods, the SSN shows a decreasing trend, while the cosmic ray flux shows an increasing trend. The investigation further shows that in each 10-year slot during 1964-2018, both showed the opposite trends (results not shown). Besides, in every slot, the crest of one superimposes on the trough of the other (results not shown).

Thus, Figure 5 *a* and *b* show that the SSN has a decreasing trend, while the cosmic ray flux has an increasing trend in recent years. It implies that the chance of occurrence of viral diseases in very high in the year 2019-2020. Table 1 also shows that the SSN minima are always associated with the outbreak of viral diseases.

The next question the authors wish to address is that for how long the Corona will keep creating havoc. For this purpose, the authors have investigated the monthly trends of the SSN in the recent 10-year slot, i.e. during 2010-2019. The investigation shows that in every month during 2000-2019, the SSN shows a decreasing trend. The results of February, March, and April are respectively shown in Figure 6 *a-c*. Results for other months are not shown. Thus, the study shows that every month in 2020 will show the SSN low in the current solar cycle. Thus, the authors conclude that the Corona will be present on the Earth during the whole of 2020. Figure 7 shows the monthly mean SSN from January 2019- February 2020. Figure 7 reveals that there is a decreasing trend in SSN in this period. Thus, the SSN is likely to decrease further in the coming months from March- December 2020. The situation may improve most probably in 2021, or 2022 when the SSN will gradually increase. The Earth will again witness an attack of a virus, maybe a new one with a different mutation in its DNA different from the Corona, flu, cholera or whatever.

Now the question is why the outbreaks originate often in China. The reason is that the thickness of the stratosphere is very low over China. The height of the tropopause in the deep equatorial region starts at about 17.3 km and near 30 degrees in the northern and southern hemisphere, it comes down at 17.1 km[32]. In the northern polar region between 60N-90N, it comes down at about 10.5 km, while in the southern polar region it comes farther down to 9.7 km[32]. The height of the stratopause is about 50.0 km in the tropical region. Near 30 degrees both in the northern and the southern hemisphere, the height of the stratopause is the lowest-at about 47.0-48.0 km. In the southern polar latitudes, the stratopause reaches an altitude of 50.0 km, while at the northern polar latitudes, it reaches as high as 55.0 km. Thus, it is realized that the thickness of the stratosphere is the minimum (about 29.9-30.9 km) near 30 degrees North and South. The thickness of the stratopause is about 32.7 km in the tropical region. At the north pole it is about 44.5 km thick, while at the South pole, it is about 40.3 km thick. Because the stratospheric thickness is the minimum near 30N and 30S, it is easy for the cosmic ray and other electrically charged viruses[20] to enter the Earth at these latitudes. Wuhan, China lies at 30.59N. Thus, the first entry of the Corona via Wuhan is scientifically reasonable. Many of the viral outbreaks had first occurred in China as the stratospheric thickness is the minimum there[20].

## Conclusion

The results of the study indicate that whenever the Sun will reach the most quiescent state, as marked by the minimum sunspot number, outbreaks of viral diseases will happen in the world. Thus, the authors would like to draw the attention of the governments of all countries, the policymakers, the astrobiologists, and the virologists to adopt necessary mitigation techniques against the unforeseen natural calamities. As the SSN minima occur after every 11-13 years, the outbreak of viral diseases is also likely to recur after every 11-13 years. So, the authors wish to state that the next catastrophic viral attack will occur again anytime in 2030-2032. The study also indicates that places near 30N and 30S are the entry points of cosmic rays to the Earth, implying that the viral outbreaks are likely to occur at first near these latitudes.

Finally, the authors conclude that the outbreak of viral diseases is a natural calamity. Probably the human race cannot stop the entry or generation of viruses. What they can do is to adopt mitigation techniques to reduce casualties.

*Conflict of interest:* The authors declare that they have no conflict of interest.

ACKNOWLEDGEMENTS

The authors wish to convey a deep sense of gratitude to the Management of Sona College of Technology, Salem, India for encouraging to carry out the work. The authors wish to express sincere gratitude to the scientific inquisitiveness of mankind that has led to the research finding. The authors wish to thank the Solar Influences Data Analysis Centre (SIDC), Royal Observatory of Belgium for the sunspot number data and the Cosmic Ray Station, Sodankyla Geophysical Observatory, University of Oulu, and the World Data Centre for Cosmic Ray, Physical Research Laboratory, Ahmedabad for the cosmic ray flux data.

**Table 1.** Occurrence of sunspot minima and the outbreak of viral diseases

| Year | Sunspot number | Year of outbreak | Location | Disease |
|---|---|---|---|---|
| 1700 | 5.0 | 1702-1703 | Canada, New France | Smallpox |
| 1711 | 0 | 1710-1712 | Denmark, Sweden | Plague |
| 1712 | 0 | 1713-1715 | Thirteen colonies | Measles |
| 1723 | 11.0 | 1723 | Thirteen Colonies | Smallpox |
| 1733 | 5.0 | 1732-1733 | Canada, New France | Smallpox |
| 1744 | 5.0 | 1743 | Italy | Plague |
| 1755 | 9.6 | 1755-1756 | North America | Smallpox |
| 1766 | 11.4 | | | |
| 1775 | 7.0 | 1775-1776 | England | Influenza |
| 1784 | 10.2 | | | |
| 1788 | 130.9 | 1788 | United States | Measles |
| 1798 | 4.1 | 1793-1798 | United States | Yellow fever |
| | | 1800-1803 | Spain | Yellow fever |
| | | 1801 | Egypt | Plague |
| 1809 | 2.5 | 1812-1819 | Ottoman Empire | Plague |
| 1810 | 0 | | | |
| 1811 | 1.4 | | | |
| 1812 | 5.0 | 1812 | Egypt | Plague |
| | | 1813-1814 | Malta (epidemic) | |
| | | 1813 | Romania | |
| | | 1816-1826 | Asia, Europe (pandemic) | Cholera |
| 1823 | 1.8 | 1820-1823 | United States | Yellow fever |
| | | 1821 | Spain | |
| 1833 | 8.5 | 1829-1835 | Iran (pandemic) | Plague |
| | | 1829-1851 | Asia, Europe, North America (pandemic) | Cholera |
| 1837 | 138.3 | 1837-1838 | Great Plains (epidemic) | Smallpox |
| 1843 | 10.7 | 1847-1848 | Canada (epidemic) | Typhus |
| 1848 | 124.7 | 1847-1848 | Worldwide | Influenza |
| 1856 | 4.3 | 1856-1860 | Worldwide (pandemic) | Plague |
| | | 1852-1860 | Russia (pandemic) | Cholera |
| | | 1863-1879 | Middle East (pandemic) | Cholera |
| 1867 | 7.3 | 1867 | Iraq | Plague |
| | | | Australia | Measles |
| 1878 | 3.4 | 1881-1896 | India, Germany (pandemic) | Cholera |
| 1889 | 6.3 | 1889-1890 | Worldwide (pandemic) | Influenza |
| 1902 | 5.0 | | | |
| 1913 | 1.4 | 1910-1912 | China | Plague |
| 1923 | 5.8 | 1915-1926 | Worldwide (pandemic) | Encephalitis |
| | | 1918-1920 | Worldwide (pandemic) | H1N1 |
| 1933 | 5.7 | 1933 | UK | Typhoid |
| 1944 | 9.6 | 1942-1944 | Egypt | Malaria |
| 1954 | 4.4 | 1957-1958 | Worldwide | Asian Flu |
| 1964 | 10.2 | 1968-1969 | Worldwide | Hong Kong flu |
| 1976 | 12.6 | 1977 | Worldwide | Smallpox |
| 1986 | 13.4 | | | |
| 1996 | 8.6 | 1996-2001 | Worldwide | VCJD |
| | | 1996 | West Africa | Meningitis |
| 2008 | 2.9 | 2008 | India | Cholera |
| 2009 | 3.1 | 2009 | Worldwide (Pandemic) | Swine flu |
| 2019 | 6.3 | 2019 | Worldwide (pandemic) | Corona |
| February 2020 | 0.4 | | Worldwide | Corona |

**Figure legend**

**Figure 1.** Variation of SSN (1700- 2019) and $^{10}$Be (1700-1994).

**Figure 2 *a*.** Yearly variation of SSN and cosmic ray flux over Oulu, Finland (1964-2018).

**Figure 2 *b*.** Yearly variation of SSN and cosmic ray flux over Ahmedabad, India (1968-1973).

**Figure 3 *a*.** Correlation between SSN and cosmic ray flux over Oulu, Finland.

**Figure 3 *b*.** Correlation between SSN and cosmic ray flux over Ahmedabad, India.

**Figure 4 *a*.** Yearly variation of cosmic ray flux and $^{10}$Be over Oulu, Finland (1964-1994).

**Figure 4 *b*.** Yearly variation of cosmic ray flux and $^{10}$Be over Ahmedabad, India (1968-1973).

**Figure 5 *a*.** Variation of SSN and cosmic ray flux over Oulu, Finland (1979-1998).

**Figure 5 *b*.** Variation of SSN and cosmic ray flux over Oulu, Finland (1999-2018).

**Figure 6 *a*.** Monthly trend of SSN in February (2010-2020).

**Figure 6 *b*.** Monthly trend of SSN in March (2010-2019).

**Figure 6 *c*.** Monthly trend of SSN in April (2010-2019).

**Figure 7.** Variation of monthly mean SSN (January 2019-February 2020).

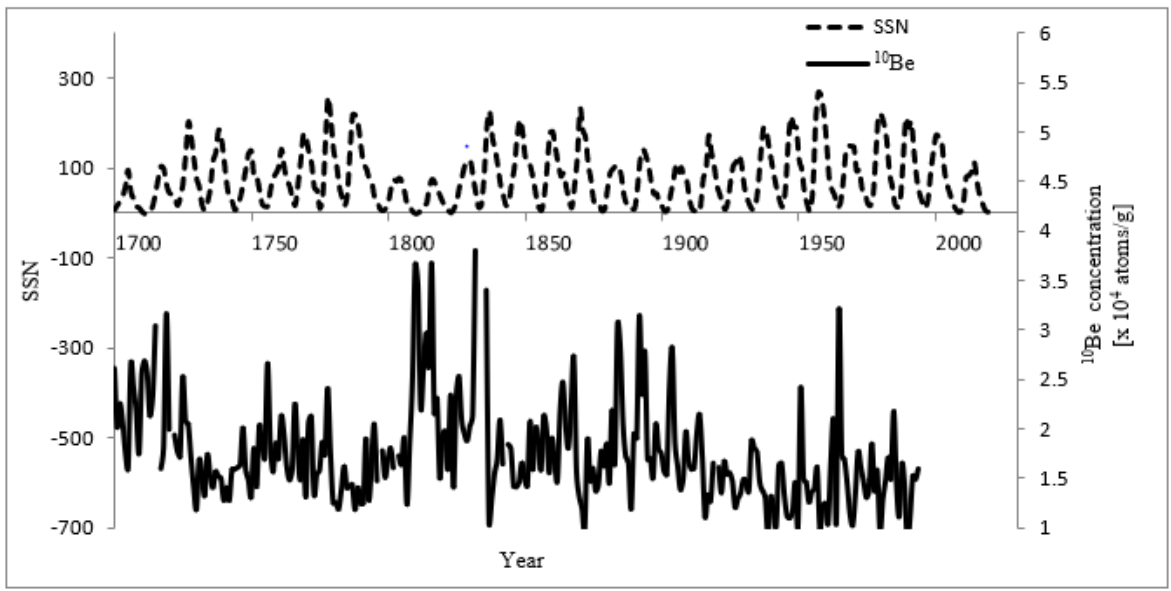

**Figure 1.** Variation of SSN (1700- 2019) and $^{10}$Be (1700-1994).

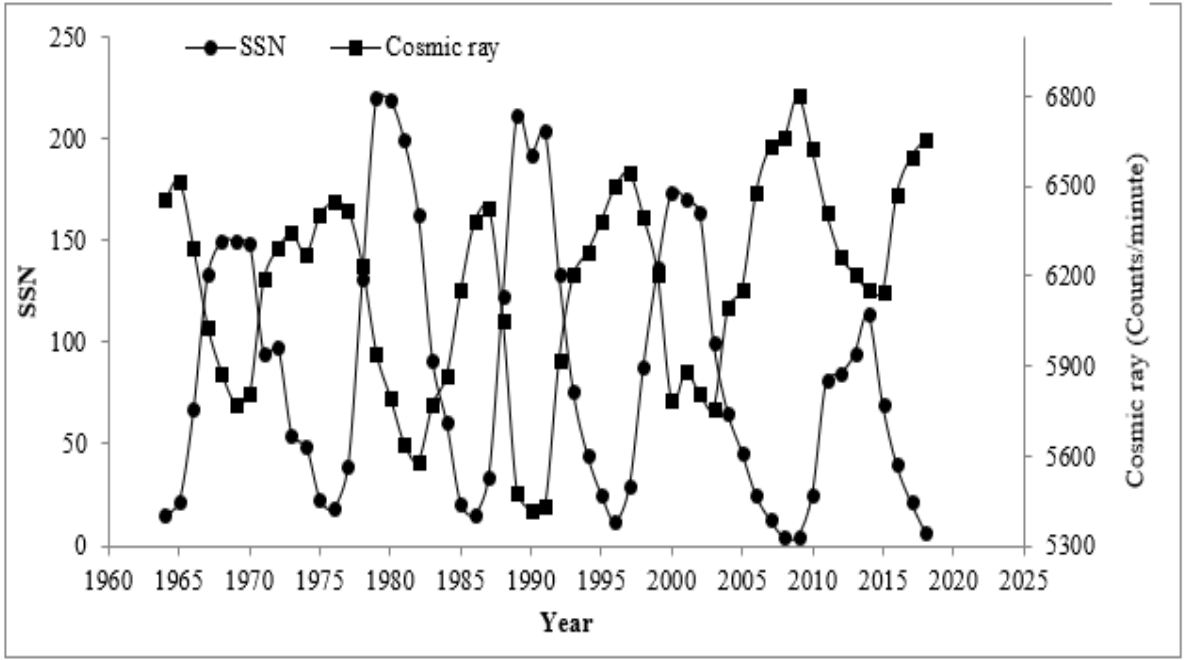

**Figure 2 *a*.** Yearly variation of SSN and cosmic ray flux over Oulu, Finland (1964-2018).

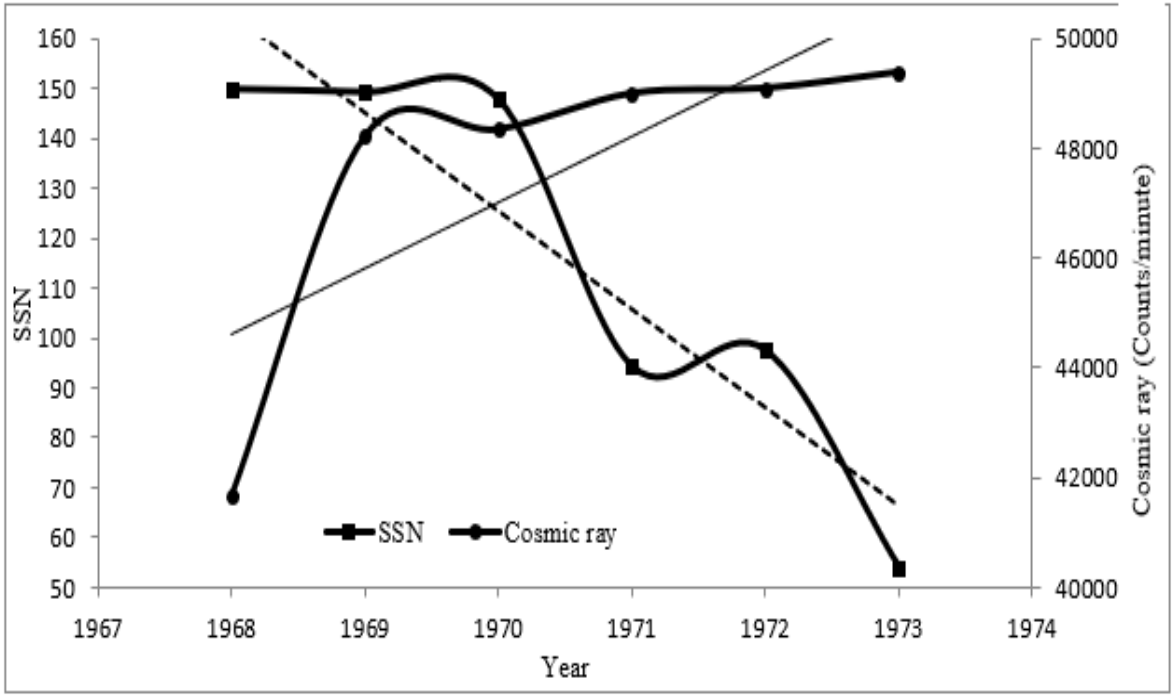

**Figure 2 *b*.** Yearly variation of SSN and cosmic ray flux over Ahmedabad, India (1968-1973).

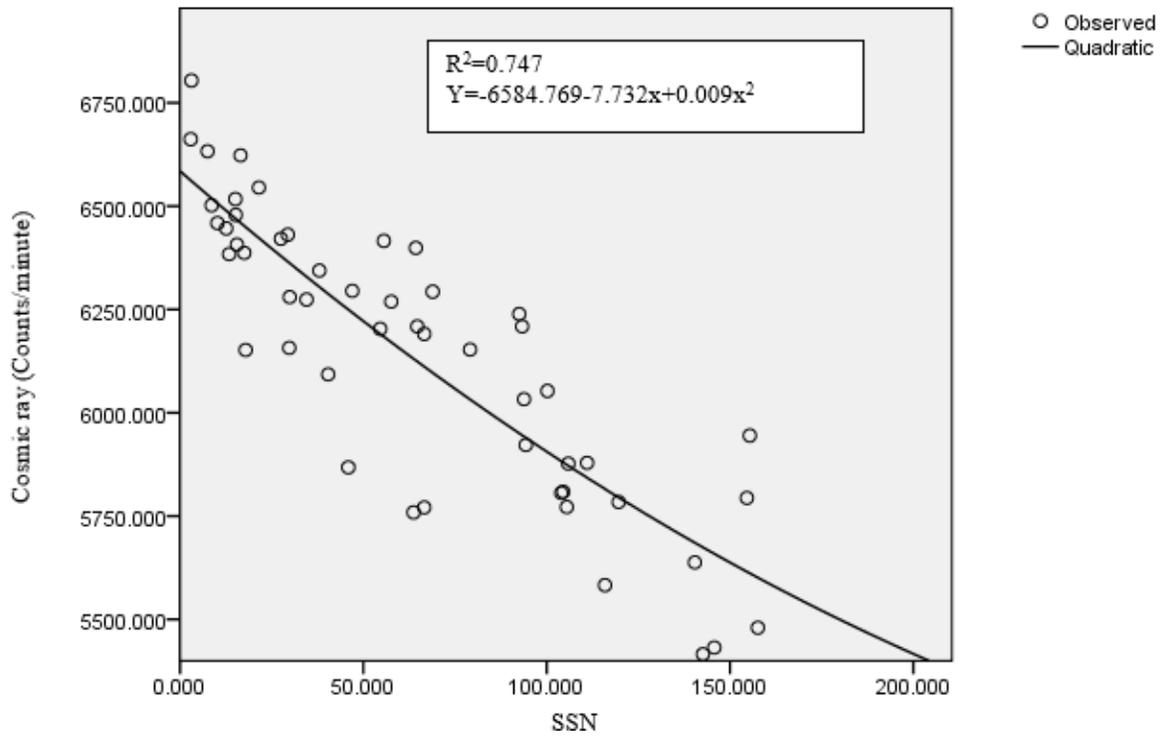

**Figure 3 *a*.** Correlation between SSN and cosmic ray flux over Oulu, Finland.

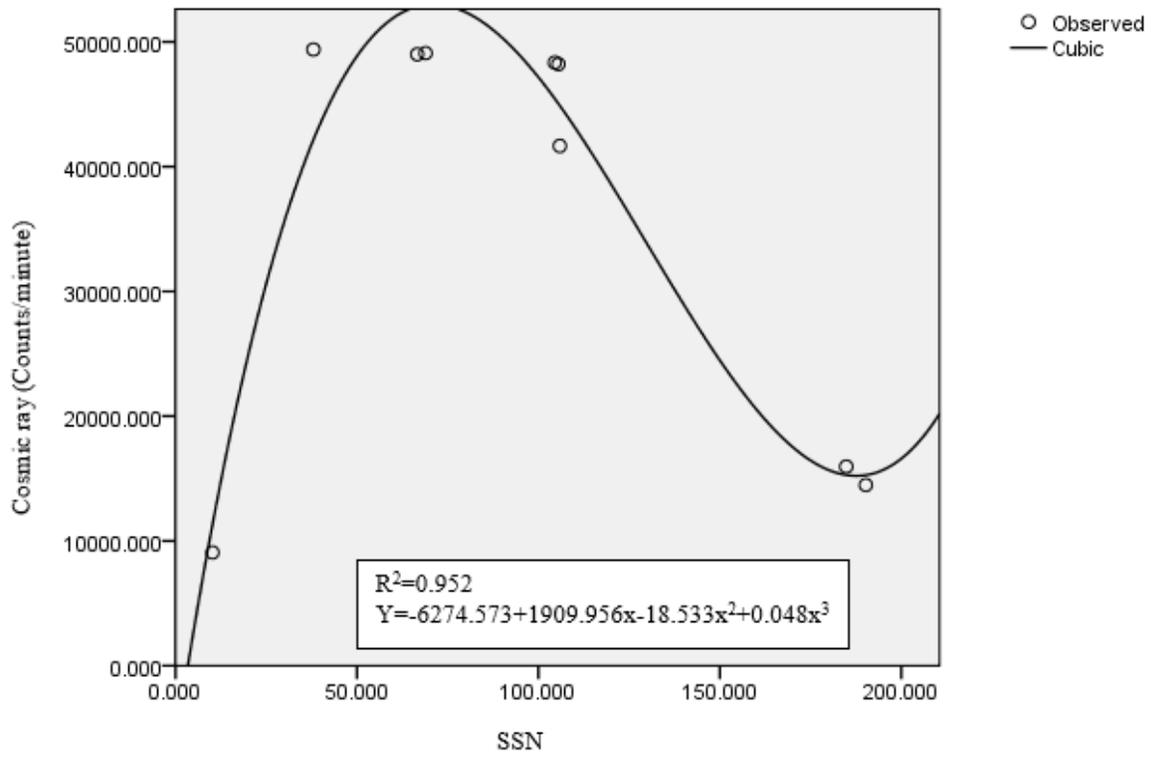

**Figure 3 *b*.** Correlation between SSN and cosmic ray flux over Ahmedabad, India.

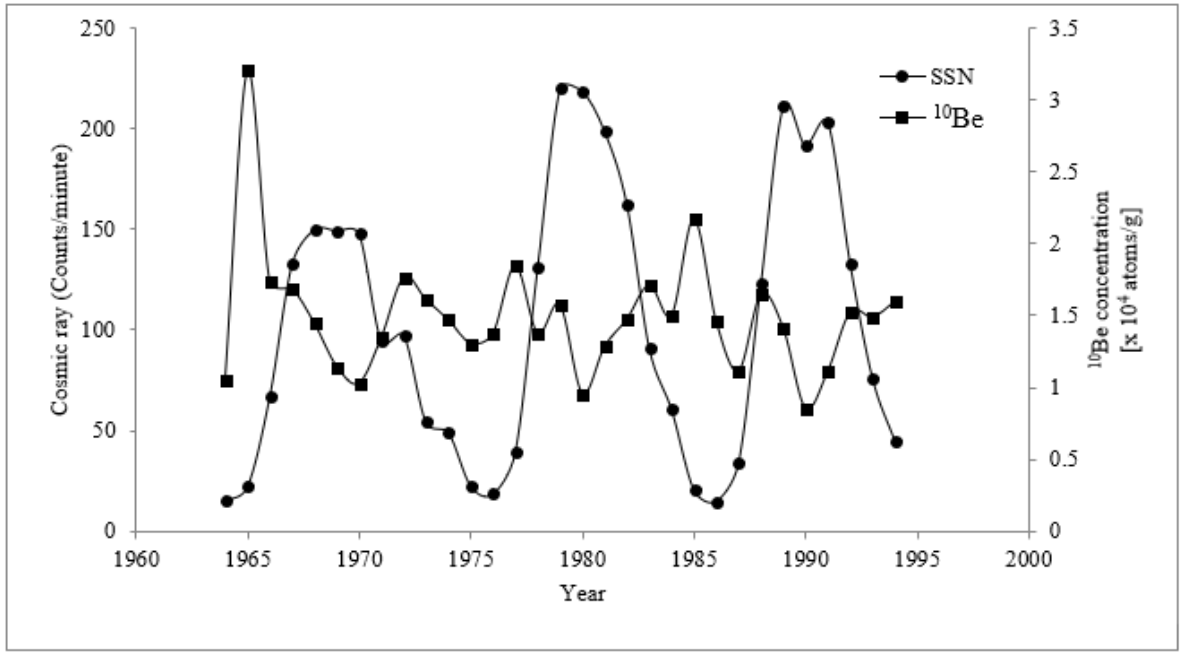

**Figure 4 *a*.** Yearly variation of cosmic ray flux and $^{10}$Be over Oulu, Finland (1964-1994).

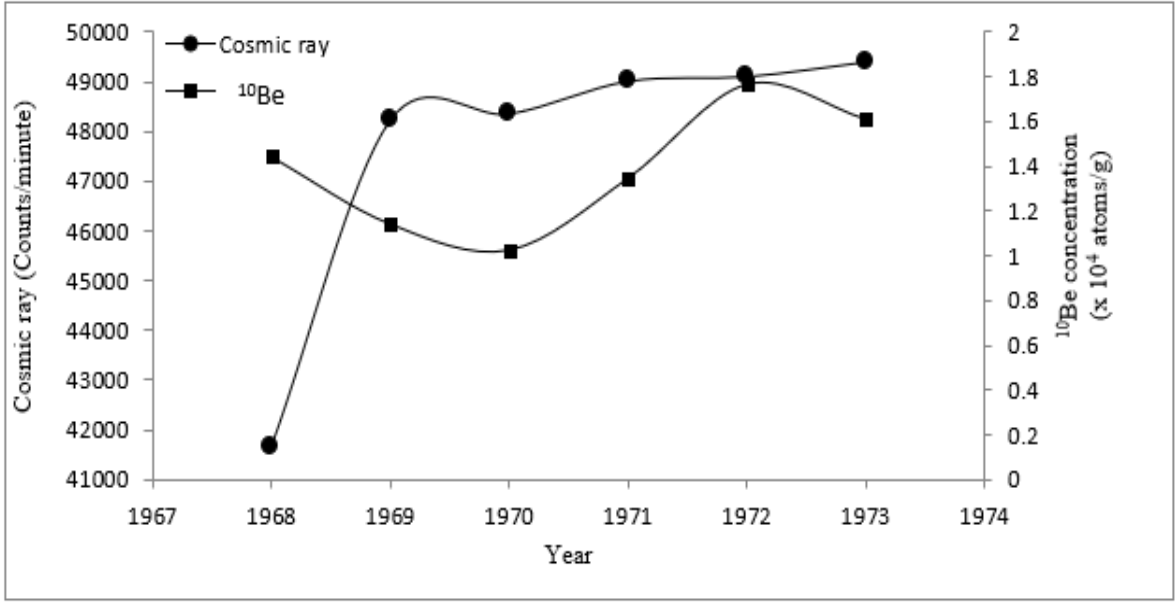

**Figure 4 *b*.** Yearly variation of cosmic ray flux and $^{10}$Be over Ahmedabad, India (1968-1973).

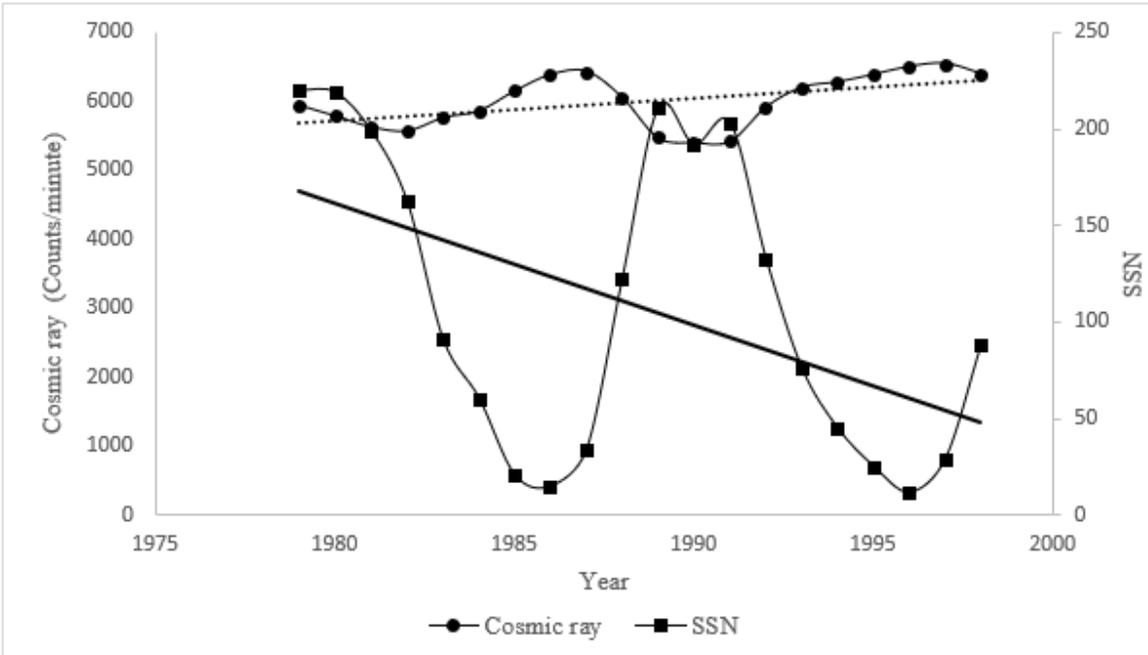

**Figure 5 *a*.** Variation of SSN and cosmic ray flux over Oulu, Finland (1979-1998).

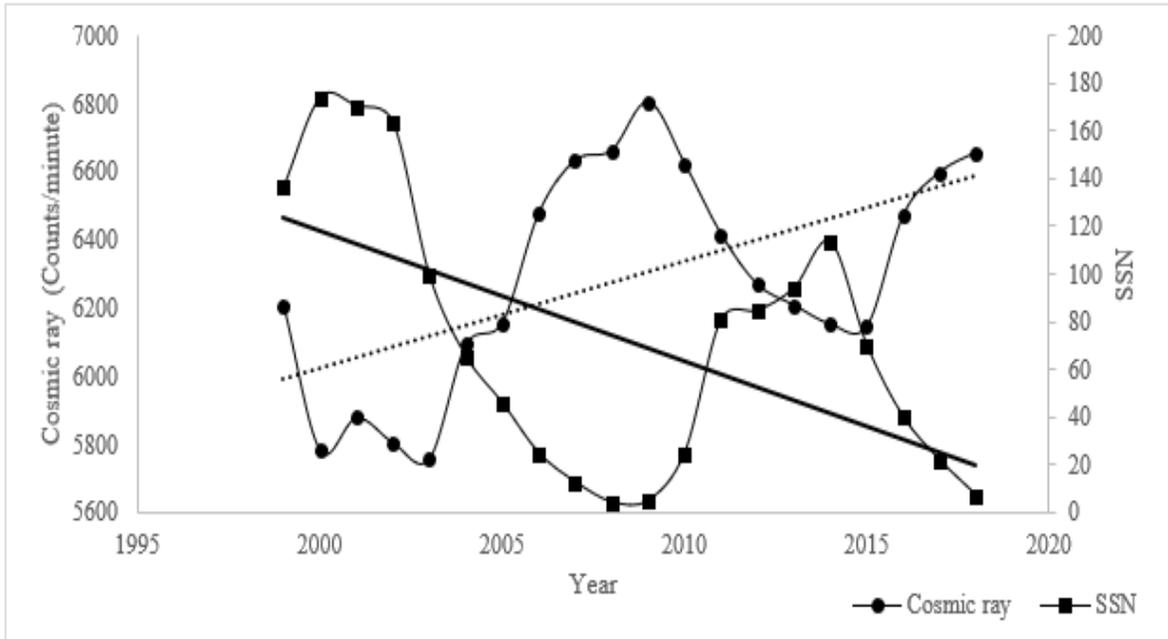

**Figure 5 *b*.** Variation of SSN and cosmic ray flux over Oulu, Finland (1999-2018).

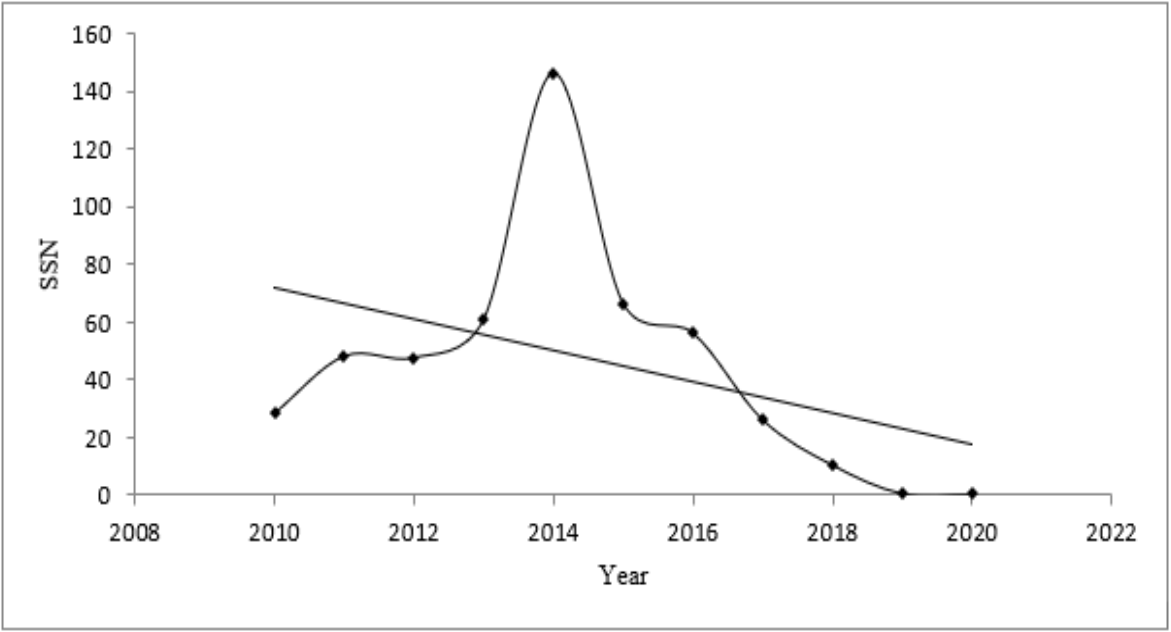

**Figure 6 *a*.** Monthly trend of SSN in February (2010-2020).

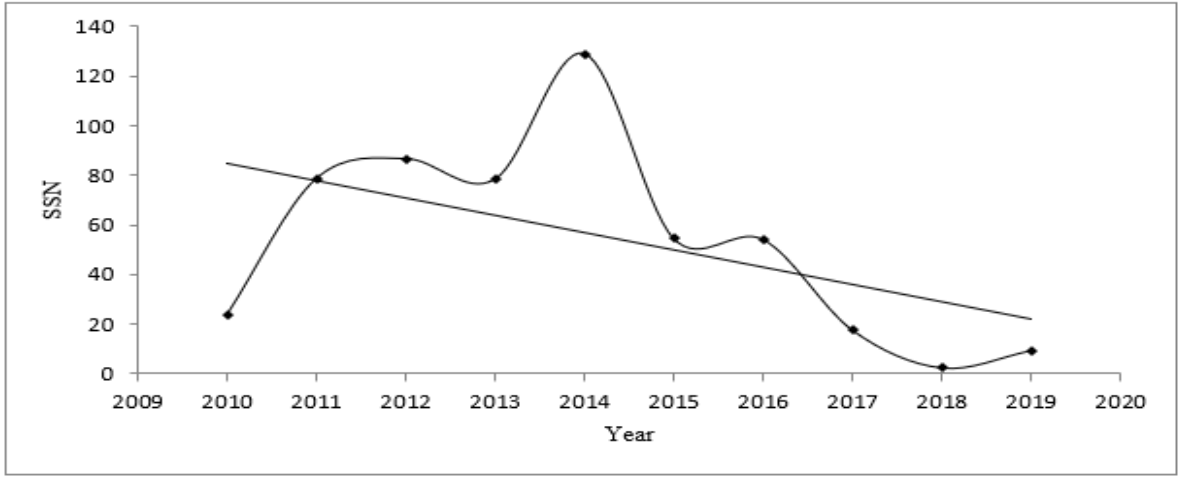

**Figure 6 *b*.** Monthly trend of SSN in March (2010-2019).

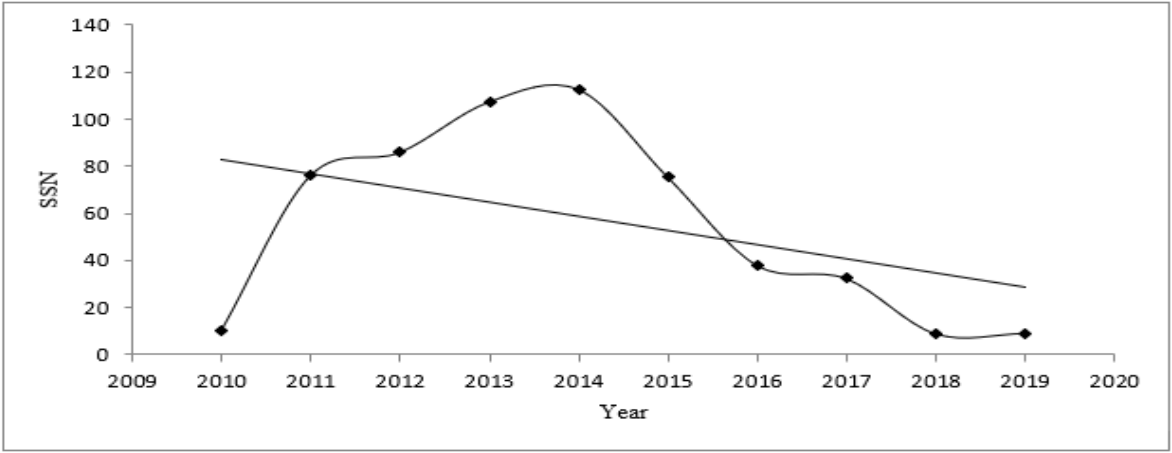

**Figure 6 *c*.** Monthly trend of SSN in April (2010-2019).

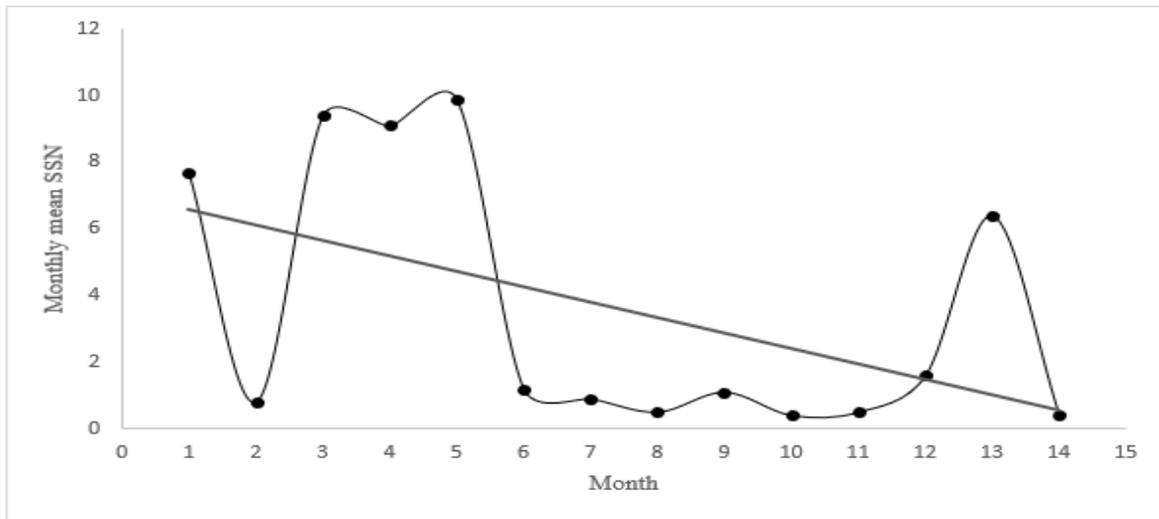

**Figure 7.** Variation of monthly mean SSN (January 2019-February 2020).